\documentclass[useAMS,usenatbib,usegraphicx]{mn2e}

\usepackage{amsmath}
\usepackage{color}
\title[Early Optical Afterglows of GRB 090426]
   {Probing the Nature of High-$z$ Short GRB 090426 with Its Early  Optical and X-ray Afterglows}

\author[Xin et al.]
{Li-Ping Xin$^{1}$\thanks{email: xlp@bao.ac.cn},
En-Wei Liang$^{2}$\thanks{email: lew@gxu.edu.cn}, Jian-Yan Wei$^{1}$, Bing Zhang$^{3}$, Hou-Jun Lv$^{2}$,
\newauthor
Wei-Kang Zheng$^{4}$,
Yuji Urata$^{5}$, Myungshin Im$^{6}$, Jing Wang$^{1}$, Yu-Lei Qiu$^{1}$,
\newauthor
Jin-Song Deng$^{1}$,
Kui-Yun Huang$^{5}$,
Jing-Yao Hu$^{1}$, Yiseul Jeon$^{6}$, Hua-Li Li$^{1}$
\newauthor
and Xu-Hui Han$^{1}$  \\
\\
$^1$ National Astronomical Observatories, Chinese Academy of Sciences, Beijing 100012, China.\\
$^2$ Department of Physics, Guangxi University, Guangxi 530004, China.\\
$^3$ Department of Physics and Astronomy, University of Nevada, Las Vegas, Nv 89154, USA.\\
$^4$ Department of Physics, University of Michigan, Ann Arbor, MI 48109, USA. \\
$^5$ Institute of Astronomy and Astrophysics, Academia Sinica, P.O. Box 23-141, Taipei 106, Taiwan.\\
$^6$ Center for the Exploration of the Origin of the Universe, Department of Physics \& Astronomy, FPRD, Seoul National University, \\
Shillim-dong, San 56-1, Kwanak-gu, Seoul, Korea.
}

\begin{document}
\label{firstpage}
\date{Accepted Received: Revision 1}
\pagerange{\pageref{firstpage}--\pageref{lastpage}}
\pubyear{2010}
\maketitle

\begin{abstract}
 GRB 090426 is a short duration burst detected by Swift ($T_{90}\sim 1.28$ s in the
observer frame, and $T_{90}\sim 0.33$ s in
the burst frame at $z=2.609$). Its host galaxy properties and some
$\gamma$-ray related correlations are analogous to those seen in long duration
GRBs, which are believed to be of a massive-star origin (so-called Type II GRBs).
We present the results of its early optical observations with the
0.8-m TNT telescope at Xinglong observatory, and the 1-m LOAO  telescope at Mt.
Lemmon Optical Astronomy Observatory in Arizona. Our well-sampled optical
afterglow lightcurve covers from $\sim 90$ seconds to $\sim 10^4$ seconds post
the GRB trigger. It shows two shallow decay episodes that are likely due to
energy injection, which end at $\sim 230$ seconds
and $\sim 7100$ seconds, respectively. The decay slopes post the injection
phases are consistent with each other ($\alpha\simeq 1.22$). The X-ray
afterglow lightcurve appears to trace the optical, although the second energy
injection phase was missed due to visibility constraints introduced by the {\em Swift} orbit.
The X-ray spectral index is $\beta_X\sim 1.0$ without temporal evolution. Its
decay slope is consistent with the prediction of the forward shock model. Both
X-ray and optical emission is consistent with being in the same spectral regime
above the cooling frequency ($\nu_c$). The fact that $\nu_c$ is below the
optical band from the very early epoch of the observation provides a constraint on
the burst environment, which is similar to that seen in classical long duration GRBs.
We therefore suggest that death of a massive star is the possible
progenitor of this short burst.
\end{abstract}


\section{Introduction}
Cosmic gamma-ray bursts (GRBs) are classified into two classes with a
separation at the observed burst duration of $T_{90}\sim 2$ seconds based on CGRO/BATSE
observations (Kouveliotou et al. 1993). The afterglow and host galaxy properties of long GRBs,
especially the detections of several long GRB-supernova
associations (e.g. Galama et al. 1998; Hjorth et al. 2003; Soderberg et al. 2004;
Campana et al. 2006), suggest that they are mostly likely related to deaths of massive
stars. The ``collapsar" model has been widely recognized as the standard
scenario for long GRBs (Woosley 1993; Paczy\'{n}ski 1998; Zhang \&
M\'{e}sz\'{a}ros 2004; Piran 2004; Woosley \& Bloom 2006).
In the {\em Swift} era, the afterglows and host galaxies of some
short GRBs have been identified (Gehrels et al. 2005;
Villasenor et al. 2005; Fox et al. 2005; Berger et al. 2005; McGlynn et al.
2008).  Some short GRBs are found to be
associated with nearby early-type galaxies with little
star formation. Some others are located in late-type star forming galaxies
(e.g. Fox et al. 2005; Fong et al. 2010), some of which are at high redshifts
(Levan et al. 2006; Berger et al. 2007).
No short-duration GRB was found to be associated with a supernova (Kann et al. 2008;
Zhang et al. 2009, and references therein). All these seem to favor the idea that
short GRBs are from mergers of two compact
stellar objects (Eichler et al. 1989; Narayan et al. 1992).

Later {\em Swift} observations have revealed that the short/long
separation is not sufficiently clean to differentiate their physical
origins. This led to introduction of the
physical classification scheme Type II
(collapses of massive stars) vs. Type I (putatively identified as mergers) GRBs
(Zhang et al. 2007a; Kann et al. 2007, 2008; Zhang et al. 2009; Lv et al. 2010). 
Some convincing Type I GRBs have long, soft ``extended emission" (Barthelmy et al.
2005; Norris et al. 2006; Lin et al. 2008; Zhang et al. 2009; Perley et al.
2009), making their $T_{90}$ ``long". The non-detection of any supernovae
associated with the nearby long GRBs 060614 and 060505 (Gehrels et al. 2006;
Gal-Yam et al. 2006; Fynbo et al. 2006) casted doubts on the Type II origin for
these long GRBs. Some long GRBs have rest-frame durations shorter than
2 seconds (Levan et al. 2007). Observations of two intrinsically
short-duration, high-$z$ GRBs 080913 ($z=6.7$; Greiner et al. 2009a) and 090423
($z=8.3$; Tanvir et al. 2009; Salvaterra et al. 2009) suggest that they share a
lot of common properties with long GRBs, and most likely have a massive star
progenitor (Zhang et al. 2009; Lin et al. 2009; Levesque et al. 2010a;
Belczynski et al. 2010). Zhang et al. (2009) and Virgili et al. (2009)
suggested that some (or even most) short duration GRBs are probably not
produced via compact star mergers (Type I), but are likely related to massive
stars (Type II). Based on the observed gamma-ray energy and peak energy of the
$\nu f_\nu$ spectrum of prompt gamma-ray emission, Lv et al. (2010) defined a
parameter $\varepsilon \equiv E_{\rm iso}/E_{p,z}^{1.7}$, and proposed a new
empirical classification scheme that is found to better match the
physically-motivated Type II/I classification scheme. They showed that the
typical Type II GRBs are in the high-$\varepsilon$ group, in contrast to the
typical type I GRBs, which belong to the low-$\varepsilon$ group.

Another striking case that poses a challenge to the conventional long vs. short
GRB classification scheme is GRB 090426. This burst has an observed $T_{90}=
1.29\pm 0.09$ seconds in the {\em Swift} BAT band (Sato et al. 2009), corresponding
to a burst rest-frame duration of 0.33 s at redshift $z=2.609$ (Levesque et al.
2010a). Phenomenologically, in terms of duration alone,
it is unambiguously within the range of classical
short-type GRBs in both the observer frame and the burst rest frame. On the other
hand, both the host galaxy properties (Levesque et al. 2010a) and the spectral energy
properties (Antonelli et al. 2009) suggest that it is mostly likely a Type II GRB,
i.e., related to core collapse of a massive star. With the new classification
method proposed by Lv et al. (2010), this event is also well classified into
the high-$\varepsilon$ group, which is where all known Type II GRBs belongs to.

Multi-wavelength afterglows are essential for revealing the burst environment,
and hence can serve as a probe of the GRB progenitor. In this paper, we
report our observations of the early optical afterglow for GRB 090426 using
the TNT telescope at Xinglong Observatory and the LOAO telescope at Mt. Lemmon
Optical Astronomy Observatory in Arizona. We use the optical and X-ray afterglow
data to explore the nature of this event.
Our observations are reported in Section 2. A joint optical and X-ray
data analysis is present in Section 3. Conclusions and discussion
are present in Section 4. The notation $f_\nu\propto t^{-\alpha}\nu^{-\beta}$ is
used throughout the paper, where $f_\nu$ is the spectral flux density at the
frequency $\nu$.

\section{Observations}
GRB 090416 was detected by {\em Swift} Burst Alert Telescope (BAT) at 12:48:47
UT on 2009 April 26 (Cummings et al. 2009). Its duration is $T_{90}=1.28\pm
0.09$ sec in 15-350 KeV. The {\em Swift} X-Ray Telescope (XRT) began to observe
the burst since 84.6 sec after the GRB trigger. At 89 sec after the
trigger, The {\em Swift} UVOT  began to observe
the burst and reported an optical counterpart with a brightness
of about 17.5 mag in the white band. The optical afterglow was confirmed by
the Xinglong TNT telescope (Xin et al. 2009) and other follow-up observations
(e.g. Im et al. 2009). A redshift of 2.609 was determined by Levesque et al.
(2009c) using the Keck telescope, which was confirmed by the ESO VLT
observation (Thoene et al. 2009). The time-integrated $\gamma$-ray spectrum is
well fit by a power-law with a photon index $\Gamma$ of 1.93 (Sato et al. 2009)
($\beta=\Gamma-1$), which
roughly corresponds to an estimated spectral peak energy (in the observer frame)
$E_{p,obs}\sim 45$ keV using an empirical relation between $E_p$ and the power-law
photon index of the BAT spectrum (Zhang et al. 2007a; Sakamoto et al. 2009).

\subsection{Optical Observation and Data Reduction}
We carried out a follow-up observation campaign of GRB 090426 using
the TNT (0.8-m Tsinghua University - National Astronomical Observatory of China
Telescope) at Xinglong Observatory, under the framework of East-Asia GRB
Follow-up Observation Network (EAFON, Urata et al. 2003; 2005). TNT is
equipped with a PI $1300\times1340$ CCD and filters in the standard Johnson
Bessel system. Its field of view is $11.4\times11.4$ arcmin, yielding a 0.5
arcsec pixel scale. A custom-designed automation system has been
developed for the GRB follow-up observations (Zheng et al. 2008).

The observation of the optical transient (OT) of GRB 090426 was carried out with
TNT at 86 seconds post the {\em Swift}/BAT trigger, which is slightly earlier
than the beginning observation of XRT and UVOT on-board {\em Swift}.
A new fading source was discovered and confirmed as the OT of the burst
(Xin et al. 2009). The coordinates of the OT are consistent with that of UVOT
(Cummings et al. 2009). The $W$ (white), $R$ and $V$-band images were obtained in
86$-$519 seconds, $570-1513$ seconds, and $1907-10748$ seconds post the GRB
trigger, respectively.

The 1-m  telescope (Han et al. 2005) is located at Mt. Lemmon Optical Astronomy
Observatory LOAO in Arizona operated by the
Korea Astronomy Space Science Institute. The observation of GRB 090426 was carried
out at about 16.3 hours after the burst ($t\sim 60$ ks) in the $R$ filter.
The OT was not detected and only upper limits were obtained.

Data reduction was carried out following the standard routine in
IRAF\footnote{IRAF is distributed by NOAO, which is operated by AURA, Inc., under
cooperative agreement with NSF.} package, including bias and flat-field
corrections. Dark correction was not performed since the temperature of our CCD
was cooled down to $-110\,^{\circ}\mathrm{C}$. Point spread function (PSF)
photometry was applied via the DAOPHOT task in the IRAF package to obtain the
instrumental magnitudes. During the reduction, some frames were combined in
order to increase the signal-to-noise ratio (S/N). In the calibration and analysis,
the white band was treated as the $R$ band (Xin et al. 2010).
Absolute calibration was performed using the Sloan Digital Sky Survey (SDSS,
Adelman-McCarthy et al. 2008), with conversion of SDSS to Johnson-Cousins
system\footnote{http://www.sdss.org/dr6/algorithms/sdssUBVRITransform.html
\#Lupton2005}. The data of GRB 090426 obtained by TNT and LOAO are reported in
Table.~\ref{Tab:publ-data}.
%
\begin{table}
\caption[Optical Afterglow Photometry Log of GRB 090426]{Optical Afterglow
Photometry Log of GRB 090426. The
reference time $T_0$ is {\em Swift} BAT burst trigger time. All Data are not
corrected for the Galactic extinction (which is $E_{B-V}=0.017$, Schlegel et al.1998).
``Merr" means the uncertainty of magnitude.}
  \label{Tab:publ-data}
  \begin{center}\begin{tabular}{cccccc}
  \hline\noalign{\smallskip}
T-T0(mid) &  Exposure & Mag & Merr & Filter & Telescope \\
   sec        &       sec   &    &       &        &           \\
  \hline\noalign{\smallskip}
86    &  20.0          &  16.44  &  0.03  &  W  &  TNT  \\
109   &  20.0          &  16.50  &  0.03  &  W  &  TNT  \\
133   &  20.0          &  16.59  &  0.03  &  W  &  TNT  \\
155   &  20.0          &  16.66  &  0.03  &  W  &  TNT  \\
178   &  20.0          &  16.76  &  0.03  &  W  &  TNT  \\
201   &  20.0          &  16.88  &  0.04  &  W  &  TNT  \\
223   &  20.0          &  17.01  &  0.04  &  W  &  TNT  \\
246   &  20.0          &  17.01  &  0.04  &  W  &  TNT  \\
269   &  20.0          &  17.07  &  0.04  &  W  &  TNT  \\
292   &  20.0          &  17.09  &  0.04  &  W  &  TNT  \\
314   &  20.0          &  17.22  &  0.04  &  W  &  TNT  \\
337   &  20.0          &  17.30  &  0.05  &  W  &  TNT  \\
360   &  20.0          &  17.31  &  0.05  &  W  &  TNT  \\
382   &  20.0          &  17.44  &  0.05  &  W  &  TNT  \\
405   &  20.0          &  17.50  &  0.06  &  W  &  TNT  \\
428   &  20.0          &  17.56  &  0.06  &  W  &  TNT  \\
451   &  20.0          &  17.69  &  0.06  &  W  &  TNT  \\
473   &  20.0          &  17.63  &  0.06  &  W  &  TNT  \\
496   &  20.0          &  17.80  &  0.06  &  W  &  TNT  \\
519   &  20.0          &  17.89  &  0.06  &  W  &  TNT  \\
570   &  60.0          &  17.93  &  0.06  &  R  &  TNT  \\
649   &  60.0          &  18.12  &  0.07  &  R  &  TNT  \\
728   &  60.0          &  18.30  &  0.07  &  R  &  TNT  \\
807   &  60.0          &  18.38  &  0.08  &  R  &  TNT  \\
885   &  60.0          &  18.46  &  0.09  &  R  &  TNT  \\
964   &  60.0          &  18.72  &  0.08  &  R  &  TNT  \\
1042  &  60.0          &  18.71  &  0.08  &  R  &  TNT  \\
1121  &  60.0          &  18.78  &  0.11  &  R  &  TNT  \\
1199  &  60.0          &  18.85  &  0.11  &  R  &  TNT  \\
1278  &  60.0          &  19.01  &  0.11  &  R  &  TNT  \\
1356  &  60.0          &  19.04  &  0.12  &  R  &  TNT  \\
1435  &  60.0          &  19.12  &  0.12  &  R  &  TNT  \\
1513  &  60.0          &  19.06  &  0.12  &  R  &  TNT  \\
1907  &  300$\times$2  &  19.72  &  0.12  &  V  &  TNT  \\
2542  &  300$\times$2  &  20.04  &  0.12  &  V  &  TNT  \\
3178  &  300$\times$2  &  20.10  &  0.12  &  V  &  TNT  \\
3813  &  300$\times$2  &  20.05  &  0.12  &  V  &  TNT  \\
4448  &  300$\times$2  &  20.17  &  0.12  &  V  &  TNT  \\
5084  &  300$\times$2  &  20.19  &  0.14  &  V  &  TNT  \\
5878  &  300$\times$3  &  20.30  &  0.14  &  V  &  TNT  \\
7048  &  600$\times$2  &  20.47  &  0.16  &  V  &  TNT  \\
8282  &  600$\times$2  &  20.69  &  0.17  &  V  &  TNT  \\
9515  &  600$\times$2  &  20.65  &  0.17  &  V  &  TNT  \\
10748  &  600$\times$2  &  20.71  &  0.18  &  V  &  TNT  \\
58591 &  180$\times$10 &  $>$21.67 &      &  R  & LOAO \\

  \noalign{\smallskip}\hline
  \end{tabular}\end{center}
\end{table}

\subsection{Swift/XRT X-ray Afterglow Data Reduction}
The {\em Swift}/XRT lightcurve and spectrum are extracted from the UK Swift
Science Data Centre at the University of Leicester (Evans et al.
2009)\footnote{http://www.swift.ac.uk/results.shtml}. We fit the X-ray spectrum
with the $Xspec$ package.
The time-integrated X-ray spectrum is well fit by an absorbed
power-law model, with a photon power-law index $\Gamma=2.00\pm 0.06$ from
the PC mode data. No significant host $N_H$ excess over the Galactic value
is detected.

\section {Optical and X-ray Afterglow Joint Analysis}
\subsection{Temporal Analysis}
\begin{figure}
\centering
\includegraphics[angle=0,width=0.5\textwidth]{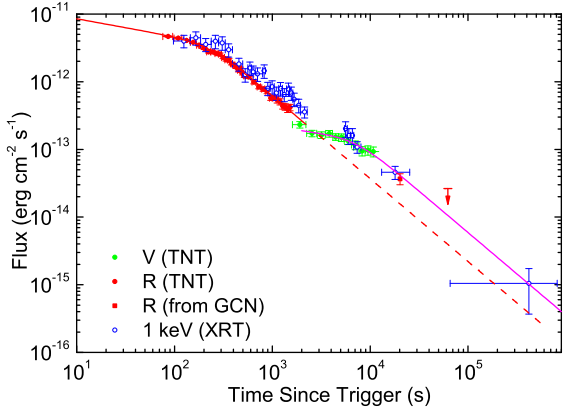}
\caption{ The extinction-corrected optical light curve of GRB 090426 (solid
dots) and the best fit with a smooth broken power-law model (solid line) for
two epoches before and after 2200 seconds post the GRB trigger. The $R$-band
data observed by Rumyantsev et al. (2009) is marked with squares. The XRT
lightcurve (open dots) is also shown for comparison. } \label{Fig:LC}
\end{figure}

With the X-ray spectral index ($\beta_X=1.00\pm 0.06$), we first derive the
1 keV lightcurve from the XRT data. It is shown in Fig.~\ref{Fig:LC}. Next, we
convert the extinction-corrected magnitudes of the optical afterglow into
energy fluxes. Levesque et al. (2010a) reported $A_{\rm V}\sim 0.4$ for the GRB
host galaxy,  and the extinction by the MilkyWay Galaxy in the burst direction is
$E_{B-V}=0.017$ (Schlegel et al.1998). For the host galaxy, the transformation
from $A_{\rm V}$ to $A_{R}$ is nearly independent of any known type of
extinction laws (MW, LMC, SMC), and we take $A_{R}\sim $ 0.32 for the host
galaxy. The Milky Way extinction corresponds to $A_{R}=0.046$ and $A_{V}=0.057$.
After extinction corrections, we find that the $\nu F_\nu$
fluxes in both the optical and the X-ray bands are almost the same.
This indicates a flat $\nu F_\nu$ spectrum from the optical to the X-ray bands,
i.e., $\beta_{OX}\sim 1.0$. This is consistent with the observed X-ray spectral
index. This flat $\nu F_\nu$ spectrum also means that we do not need to calibrate
the observed optical fluxes to a given band. The extinction-corrected optical
energy flux lightcurve is also shown in Fig.~\ref{Fig:LC}.

\begin{table}
\caption{Optical lightcurve fits with a smooth broken power-law model for two
epoches.}
  \label{Tab:fit}
  \begin{tabular}{ccccccc}
  \hline\noalign{\smallskip}
Interval  & $F_0^{*}$ &  $t_p$(s)   & $\alpha_1$ & $\alpha_2$ \\
   \hline\noalign{\smallskip}


86-2200s & 3.74$\pm$0.32 & 227$\pm$27 & 0.26$\pm$0.07 &
1.22$\pm$0.04\\

$>2200$ s &  1.77$\pm$0.67& 7123$\pm$3620 & 0.19$\pm$0.41 &
$\sim 1.22$ \\

  \noalign{\smallskip}\hline
  \end{tabular}
$^*$In units of $\times$10$^{-12}$ erg cm$^{-2}$ s$^{-1}$.
\end{table}

As shown in Fig.~\ref{Fig:LC}, the afterglow lightcurves in the optical and
X-ray bands trace with each other, suggesting that not only they are of the
similar (external shock) origin, but  they also  belong to the same
spectral regime. At early epochs  ($t<2200$ seconds post the GRB
trigger), the optical and X-ray afterglow lightcurves show a smooth
shallow-to-normal transition in the decaying behavior, with small flickerings
in the X-ray band.
In the time interval $2200 - 5500$ seconds, the brightness of the OT was
almost  constant, fading again after $t>5500$ seconds. The OT was also
detected by Rumyantsev et al. (2009) at 0.2342 days (20200 seconds) post the GRB
trigger. It had faded to $R=21.3\pm 0.2$, indicating a decay slope of $\sim
1.2$ in the time interval of $t>5500$ seconds. Due to the orbital constraints
of the {\em Swift} satellite, there is no XRT observation in the time interval
[2200, 5500] seconds when the optical lightcurve features a plateau. The fading
behavior of X-rays in the time interval [5500, 7700] seconds is similar to that
of the optical afterglow. At $t\sim 0.2342$ day, the detected X-ray behavior is
also consistent with that of the optical emission. These results suggest that
the temporal behavior of both optical and X-ray afterglows could be the same.
We thus perform a temporal analysis on the well-sampled optical lightcurve only.
The optical lightcurves in the time intervals both before and after 2200 seconds
can be well fit with a smooth broken power-law,
\begin{equation} F=F_0\left [
\left (   \frac{t}{t_b}\right)^{\omega\alpha_1}+\left (
\frac{t}{t_b}\right)^{\omega\alpha_2}\right]^{-1/\omega}.
\end{equation}
Our best-fit parameters are summarized in Table.~\ref{Tab:fit}. We find that the
decay slopes post the two breaks are similar, with a value $\sim 1.22$. The late
time X-ray data at $\sim 4\times 10^{5}$ s is also consistent with the predicted
behavior of such a decay slope in the X-ray band.

\subsection{Data Confronted with the Forward Shock Models}
As shown above, the temporal behaviors of the optical and X-ray data are consistent
with being achromatic, and the decay slopes after the two breaks observed in
the optical lightcurve are well consistent with the prediction of the forward
shock models. The fact that $\alpha_{o}=\alpha_{x}$ is consistent with
the forward shock model in the same spectra regime
(see also Urata et al. (2007) for a more general
discussion of the $\alpha_o-\alpha_x$ relation for the forward shock models).
We find no spectral evolution across the two breaks from the X-ray
data. The derived $\beta_X$ are $1.09\pm 0.15$ and $1.03\pm 0.10$, respectively,
for the integrated X-ray spectra at $t<2200$ s and $t>2200$ s. Inspecting the
spectral index $\beta_X$ and the temporal decay index $\alpha_X$ in the normal
decay segment, one finds good agreement between data and the forward shock model
closure relation in the spectral regime $\nu>\max (\nu_m,\  \nu_c)$, i.e.
$\alpha=(3\beta_X-1)/2=1.14\pm 0.23$, where $\nu_m$ and $\nu_c$ are the typical
and cooling frequencies of synchrotron radiation, respectively. Since both
optical and X-ray emissions are consistent with the forward shock origin in
the normal decay phase, one can naturally attribute the two shallow decay
segments to two epochs of energy injection into the blastwave (e.g., Dai \&
Lu 1998; Zhang \& Meszaros 2001; Zhang et al. 2006; Liang et al. 2007).

\subsection{Constraints on the Burst Environment}
The circumburst environment is critical to understand the nature
of a GRB. In the literature, usually two types of medium are discussed,
namely, a constant density medium relevant for interstellar medium (ISM)
or a stratified stellar wind with a density profile $n \propto r^{-2}$.
For the spectral regime identified for GRB 090426, i.e.
$\nu>\max (\nu_m,\  \nu_c)$, unfortunately the observed flux does not
depend on the medium density. Consequently, one can not distinguish the
two types of medium. However, since a wind model would undoubtedly point
towards a massive star progenitor, we only focus on the constant density
case. According to the analysis in section 3.2,
$\nu_c$ should be below the optical band at very early epochs.
This would give an interesting constraint on the medium density.

In the constant density case, the typical synchrotron emission frequency, the
cooling frequency and the peak spectral flux density are (Sari et al. 1998;
coefficients taken from Yost et al. 2003; Zhang et al. 2007b):
\begin{equation}
\nu_m = 3.3 \times 10^{12} {\rm\ Hz}
\left(\frac{p-2}{p-1}\right)^2(1+z)^{1/2}
\epsilon_{B,-2}^{1/2} \epsilon_{e,-1}^{2}E_{K,52}^{1/2} t_d^{-3/2} \label{num}
\end{equation}
\begin{equation}
\nu_c  =  6.3 \times 10^{15} {\rm\ Hz} (1+z)^{-1/2} (1+Y)^{-2}
\epsilon_{B,-2}^{-3/2}E_{K,52}^{-1/2} n^{-1} t_d^{-1/2} \label{nuc}
\end{equation}
\begin{equation}
F_{\nu,\max} =  1.6 {\rm\ mJy}
(1+z)D^{-2}_{28}\epsilon_{B,-2}^{1/2}E_{K,52}n^{1/2} \label{Fnumax}
\end{equation}
where $D$ is the luminosity distance in units
of $10^{28}$ cm, $f_p$ is a function of $p$ ($f_p\sim 1$ for $p= 2$, Zhang et al.
2007b), and $t_d$ is the observer's time in unit of days. The convention $Q_n=Q/
10^n$ is adopted in cgs units. The Inverse Compton scattering parameter
\begin{equation} Y=[-1+(1+4 \eta_1\eta_2 \epsilon_e /
\epsilon_B)^{1/2}]/2  \label{Y},
\end{equation}
where $\eta_1 = {\rm min} [1,(\nu_c/ \nu_m)^{(2-p)/2}]\sim 1$ for $p\sim 2.0$
(Sari \& Esin 2001), and $\eta_2 \leq 1$ is a correction factor introduced by
the Klein-Nishina effect. We take $\eta_2=1$ in our analysis, so that the
$R$-band energy flux reads
\begin{eqnarray}
\nu_R F_{\nu_R}& = & F_{\nu,\max}\nu_c^{1/2}\nu_m^{(p-1)/2}\nu_R^{(2-p)/2}
\nonumber \\
& = & 2.67\times 10^{-11} f_p ~{\rm ergs~s^{-1} ~cm^{-2}}
D_{28}^{-2}(1+z)^{(p+2)/4} \nonumber \\ & \times & (1+Y)^{-1}
\epsilon_{B,-2}^{(p-2)/4} \epsilon_{e,-1}^{p-1} E_{K,52}^{(p+2)/4}
t_d^{(2-3p)/4}\nu_{R}{^{(2-p)/2}}~, \label{FnuX-1}
\end{eqnarray}
where
\begin{equation}
f_p = \left[7.45\times 10^{-3} \left(\frac{p-2}{p-1}\right)^{2}\right
]^{(p-1)/2}.
\end{equation}
Since $p=2\beta=2.18\pm 0.30$, we have $f_p=6.04\times 10^{-3}$. One can then
derive $E_K$ using the data at any time $t_d$ (Zhang et al. 2007b):
\begin{eqnarray}
E_{K,52}&=&\left[\frac{\nu_R F_{\nu_{R}} }{1.61\times 10^{-13} ~{\rm
ergs~s^{-1} ~cm^{-2}} }\right]^{4/(p+2)} \nonumber
\nonumber \\
&\times& D_{28}^{8/(p+2)}(1+z)^{-1} t_d^{(3p-2)/(p+2)}
\nonumber \\
&\times&  (1+Y)^{4/(p+2)} \epsilon_{B,-2}^{(2-p)/(p+2)}
\nonumber \\
&\times& \epsilon_{e,-1}^{4(1-p)/(p+2)} \nu_{R}{^{2(p-2)/(p+2)}}. \label{EK1}
\end{eqnarray}
The analysis above is valid for $p>2$. From the X-ray data, we find the $p$ is
slightly larger than 2. In order to make a rough estimate, we take $p\sim 2$ to
simplify Eq. \ref{EK1} as
\begin{equation}
E_{k,52}=\frac{\nu_R F_{\nu_R}}{1.61\times 10^{-13} ~{\rm ergs~s^{-1}
~cm^{-2}}} \frac{t_d}{1+z} \epsilon_{e,-1}^{-1} (1+Y) D_{L,28}^{2}.
\end{equation}
The cooling frequency $\nu_c$ is given by
\begin{eqnarray}
\nu_c & = & 6.3 \times 10^{15} {\rm\ Hz} \left(\frac{\nu_R
F_{\nu_R}}{1.61\times 10^{-13} ~{\rm ergs~s^{-1}
~cm^{-2}}}\right)^{-1/2}\nonumber \\
&\times & (1+Y)^{-5/2} \epsilon_{B,-2}^{-3/2}\epsilon_{e,-1}^{1/2} n^{-1}
t_d^{-1}D_{28}^{-1} \label{nuc}.
\end{eqnarray}
At $t\sim 100$ seconds, one has $\nu_R F_{\nu_R}=4.42 \times 10^{-12}$ erg cm$^{-2}$
s$^{-1}$, so that
\begin{eqnarray}
\nu_c=1.57 \times 10^{17} {\rm\ Hz}(1+Y)^{-5/2}
\epsilon_{B,-2}^{-3/2}\epsilon_{e,-1}^{1/2} n^{-1} \label{nuc}.
\end{eqnarray}
The requirement of $\nu_R>\nu_c$ at $t\sim 100$ seconds then gives
\begin{equation}
n>354s(\epsilon_e, \epsilon_B),
\end{equation}
where $s(\epsilon_e,
\epsilon_B)=\epsilon_{B,-2}^{-3/2}\epsilon_{e,-1}^{1/2}(1+Y)^{-5/2}$.
This already points towards a large medium density.
Noticing that $s$ would decrease
when $\epsilon_B$ increases, we consider an extreme case of energy equipartition
among radiating electrons, magnetic field and baryons, i.e., $\epsilon_e=1/3$,
$\epsilon_B=1/3$. In this case, we still get $n >11.2$
cm$^{-3}$. Such a dense medium is inconsistent with the naive expectation of a
compact star merger progenitor, which tend to occur in low-density medium as
the compact binary escapes from the star forming region due to the natal kicks
during the births of the two neutron stars. The high density, on the other hand,
is consistent with that expected for a massive star that was born in a high
density star forming region such as a molecular cloud (e.g. Chevalier et al. 2004).

Recall that a stellar wind medium also suggests a massive star connection,
we therefore  conclude that the high ISM density would favor a burst
born in a region of ongoing star formation, and hence favor a massive star progenitor
(or a Type II GRB) for GRB 090426.

\section{Discussion and Conclusions}
We have presented the early optical afterglow observations of GRB 090426 with TNT
and LOAO. Our well-sampled optical afterglow lightcurve from $\sim 90$ seconds to
$\sim 10^4$ seconds post the GRB trigger seems to exhibit two energy injection
phases that ended at $\sim 230$ seconds and $\sim 7100$ seconds, respectively.
The temporal behavior of the observed X-rays is consistent with the optical
emission. We show that both the optical and  X-ray afterglows are consistent
with the forward shock model in the spectral regime $\nu>\max(\nu_m,\ \nu_c)$.
Although the medium type (ISM vs. wind) cannot be distinguished,
we can make a case of the massive star (Type II) origin of the GRB. For a constant
density medium, the required medium density is found very high (Eq.[12]),
inconsistent with the expectation of
a compact star merger (Type I) progenitor. A massive star
progenitor forming in a high-density star forming region (e.g. molecular cloud)
is consistent with the data.
If the circumburst medium is a stellar wind, then it also directly points
towards a massive star progenitor. We therefore conclude that the afterglow
data of the short GRB 09046 strongly suggests that it is a Type II GRB.

GRB 090426 is of great interest because of its short duration ($T_{90}\sim
0.33$ seconds in the burst frame) and high redshift ($z=2.609$). Its redshift
significantly exceeds the previous spectroscopically confirmed high redshifts
of short GRBs, e.g. GRB 070714B ($z=0.904$, Graham et al. 2009),
GRB 051121A ($z=0.546$, Soderberg et al. 2006b),
GRB 070429($z=0.902$,Cenko et al. 2008) and GRB 090510 ($z=0.903$, Rau et al. 2009).
Its host galaxy is a blue, luminous, and star-forming galaxy (Levesque et al. 2010a),
similar to the host galaxies of typical Type II GRBs (Levesque et al.2010b).
The spectral energy correlation is also consistent with being a Type II GRB.
Here we present a third argument in favor of its Type II origin,
based on early afterglow observation and modeling.
It is interesting to note that with the new
classification method proposed by Lv et al. (2010), GRB 090426 is well
grouped into Type II GRBs, in contrast to other short GRBs such as GRB 051221
and GRB 070714B.
We also noted that GRB 090426 shares some similar properties with two
other Type II bursts, GRB 040924 and GRB 050416A. The first was a short-duration
(1.2-1.5 sec), soft spectrum (Huang et al. 2005) and
a SN 1998bw-like supernova-associated burst (Soderberg et al. 2006a;
Wiersema et al. 2008).
The second burst also exhibited a soft spectrum and a short duration in
the rest frame (T$_{90}=2.4s$, $z=0.6528$), which is
associated with a SN 1998bw-like supernova (Soderberg et al. 2007), and
is located in a circumburst medium with a large density variation.

Finally, GRB 090426 has reinforced the conclusion that
burst duration alone cannot be used to judge the physical nature of a GRB,
as discussed by Donaghy et al. (2006) and
Zhang et al. (2007a)\footnote{Both teams proposed that some long GRBs
can be physically associated with compact star mergers, but did not discuss
that short bursts can be physical related to massive star core collapses.};
Levan et al. (2007)\footnote{These authors pointed out that some long GRBs
are rest-frame short.}, and Zhang et al. (2009)\footnote{These authors first
mentioned the possibility that some observer-frame short bursts can originate
from massive stars.}.  Multi-wavelength
observational campaigns are essential to unveil the physical nature of GRBs.

\section{acknowledgement}
The authors thank the anonymous referee for helpful suggestions and comments,
and Massimiliano De Pasquale for correcting an error in the early version. This
work made use of data supplied by the UK {\it Swift} Science Data Center at the
University of Leicester. It is partially supported by the National Natural
Science Foundation of China under grants No. 10673014, 10803008, 10873002, and
the National Basic Research Program (''973" Program) of China under Grant
2009CB824800. It is also partly supported by grants NSC 98-2112-M-008-003-MY3
(Y.U.), and by NASA NNX09AT66G, NNX10AD48G, and NSF AST-0908362 (B.Z.). E. W.
L. also acknowledges the support from the Guangxi SHI-BAI-QIAN project (Grant
2007201), the Guangxi Science Foundation (2010GXNSFC013011)£¬the program for
100 Young and Middle-aged Disciplinary Leaders in Guangxi Higher Education
Institutions, and the research foundation of Guangxi University (M30520). MI
and YS are supported by the Korea Science and Engineering Foundation (KOSEF)
grant No. 2009-0063616, funded by the Korean government (MEST). K.Y.H. was
supported by NSC-99-2112-M-001-002-MY3.

\bibliographystyle{mn}

\label{lastpage}
\end{document}